\documentclass[aip,jcp,reprint]{revtex4-1}

\usepackage{inputenc}
\inputencoding{ansinew}
\usepackage[OT1]{fontenc}
\usepackage{graphicx}
\usepackage{amsmath}
\newcommand*{\E}[1]{\cdot 10^{#1}}
\usepackage[version-1-compatibility]{siunitx}

\begin{document}


\title{Identical temperature dependence of the time scales of several  
  linear-response functions of two glass-forming liquids} 

\author{Bo Jakobsen}
\email{boj@ruc.dk}
\author{Tina Hecksher}
\author{Kristine Niss}
\author{Tage Christensen}
\author{Niels Boye Olsen}
\author{Jeppe C. Dyre}
\affiliation{ DNRF centre ``Glass and Time'', IMFUFA,
    Department of Sciences, Roskilde University, Postbox 260, DK-4000
    Roskilde, Denmark}

\date{\today}

\begin{abstract}
  The frequency-dependent dielectric constant, shear and adiabatic
  bulk moduli, longitudinal thermal expansion coefficient, and
  longitudinal specific heat have been measured for two van der Waals
  glass-forming liquids, tetramethyl-tetraphenyl-trisiloxane (DC704)
  and 5-polyphenyl-4-ether.  Within the experimental uncertainties the
  loss-peak frequencies of the measured response functions have
  identical temperature dependence over a range of temperatures, for
  which the Maxwell relaxation time varies more than nine orders of
  magnitude.  The time scales are ordered from fastest to slowest as
  follows: Shear modulus, adiabatic bulk modulus, dielectric constant,
  longitudinal thermal expansion coefficient, longitudinal specific
  heat. The ordering is discussed in light of the recent conjecture
  that van der Waals liquids are strongly correlating, i.e.,
  approximate single-parameter liquids.
\end{abstract}



\maketitle 

A liquid has several characteristic times \cite{Boon1980, Brawer1985,
  March2002, Barrat2003}. One is the Maxwell relaxation time
determining how fast stress relaxes $\tau_\text{M} \equiv
\eta/G_\infty$, where $\eta$ is the shear viscosity and $G_\infty$ the
instantaneous shear modulus \cite{Harrison1976}. Other characteristic
times are identified by writing $D=a^2/\tau_D$, in which $D$ may be
particle, heat, or the transverse momentum diffusion constant, and $a$
is of order the intermolecular distance. Further characteristic times
are the inverse loss-peak frequencies (i.e., frequencies of
maximum imaginary part) of different complex
frequency-dependent linear-response function \cite{Angell2000,Kremer2002}.  For
low-viscosity liquids like ambient water the characteristic times are
all of the same order of magnitude, in the picosecond range, and only
weakly dependent on temperature.

Supercooling a liquid increases dramatically its viscosity
\cite{Angell1995,Debenedetti1996,Donth2001,Binder2005,Dyre2006a}; most
characteristic times likewise increase dramatically. The metastable
equilibrium liquid can be cooled until the relaxation times become
$10^{10}$--$10^{15}$ times larger than for the low-viscosity liquid,
at which point the system falls out of metastable equilibrium and
forms a glass. At typical laboratory cooling rates
($\text{K}/\text{min}$) the glass transition takes place when
$\tau_\text{M}$ is of order 100 seconds
\cite{Angell1995,Debenedetti1996,Donth2001,Binder2005,Dyre2006a}.

Even though most characteristic times increase dramatically when the
liquid is cooled, they are generally not identical. Different measured
quantities and different definitions of the characteristic time scale
lead to somewhat different characteristic times. A trivial example is
the difference between the time scales of the bulk modulus and its
inverse in the frequency domain, the bulk compressibility.  More
interestingly, some time scales might have quite different temperature
dependence; this is often referred to as a \textit{decoupling} of the
corresponding microscopic processes.

Several relaxation time decouplings have been reported in the
ultraviscous liquid state above the glass transition. Significant
decoupling takes place for some glass-forming molten salts like
``CKN'' (a 60/40 mixture of ${\rm Ca(NO_3)_2}$ and ${\rm KNO_3}$
\cite{Angell1964}), where the conductivity relaxation time at the
glass transition is roughly $10^4-10^5$ times smaller than
$\tau_\text{M}$ \cite{Angell1988,Angell1991}. This reflects a
decoupling of the molecular motions, with the cations diffusing much
faster than the nitrate ions \cite{Angell1991}.  A more recent
discovery is the decoupling of translational and rotational motion in
most molecular liquids, for which one often finds that molecular
rotations are 10-100 times slower than expected from the
diffusion time
\cite{Fujara1992,Cicerone1996,Sillescu1999}. This is generally believed
to reflect dynamic heterogeneity of glass-forming liquids
\cite{Fujara1992,Cicerone1996,Sillescu1999,Ediger2000}.  Angell in
1991 suggested a scenario consisting of ``a series of decouplings
which occurs on decreasing temperature'', sort of a hierarchy. He
cautiously added, though, that ``more data are urgently needed to
decide if this represents the general case'' \cite{Angell1991}.

Some linear-response functions like the
thermoviscoelastic and shear-mechanical ones are difficult to measure
reliably for ultraviscous liquids
\cite{Christensen1995,Christensen2007,Christensen2008}. To the best of
our knowledge there are no studies of their possible decoupling. This
paper presents such data, together with conventional dielectric
data. The purpose is to establish the order of relaxation times among
the different quantities and, in particular, to investigate whether or
not they show a decoupling upon approaching the glass-transition
temperature.

\section{Experimental results}
We have measured the complex, frequency-dependent dielectric constant
$\varepsilon(\omega)$, shear modulus $G(\omega)$
\cite{Christensen1995,HecksherPHD}, adiabatic bulk modulus
$K_S(\omega)$ \cite{Christensen1994b,HecksherPHD}, and longitudinal
specific heat $c_l(\omega)$ \cite{Jakobsen2010} on two van der Waals
bonded glass-forming liquids tetramethyl-tetraphenyl-trisiloxane
({DC704}) and 5-polyphenyl-4-ether ({5PPE}) (commercial vacuum-pump
oils). For {DC704} the time-dependent longitudinal thermal expansion
coefficient $\alpha_l(t)$ \cite{NissAlpha} was also measured (see 
appendix \ref{sec:mater-meth-sect} for details). We note that both
liquids have linear-response functions that to a good approximation
obey time-temperature superposition (TTS)
\cite{Olsen2001,Jakobsen2005,HecksherPHD}, i.e., their loss-peak
shapes are temperature independent in log-log plots. Moreover, the two
liquids have only small beta relaxations, they rarely crystallize, and
they are generally very stable and reproducible --- altogether these
two liquids are very suitable for fundamental studies.

Three of the measured quantities ($\alpha_l$, $c_l$, and $K_S$) are
closely related to one complete set of independent scalar
thermoviscoelastic response functions \footnote{There are 24 different
  complex, frequency-dependent scalar thermoviscoelastic response
  functions referring to isotropic experimental conditions
  \cite{Meixner1959}. Any such linear-response function may be written
  as $\partial a(\omega)/\partial b(\omega)|_c$ where $a,b,c\in
  \{T,S,p,V\}$, where $T$ is temperature, $S$ entropy, $p$ pressure,
  and $V$ volume. The 24 linear-response functions are not
  independent. A number of sets of three independent functions can be
  selected (e.g. $c_p$, $\alpha_p$ and $\kappa_T$), where all the
  remaining functions can be expressed via the generalized Onsager and
  standard thermodynamic relations \cite{Meixner1959,Bailey2008b}.}:
$c_p$, $\alpha_p$, and $\kappa_T$ (the relations are given in appendix
\ref{sec:perf-corr-single}).  Measurements of such a complete set of
three scalar thermoviscoelastic response functions are rare, if at all
existing for any glass-forming liquid.

\begin{figure}
  \centering
  \includegraphics[]{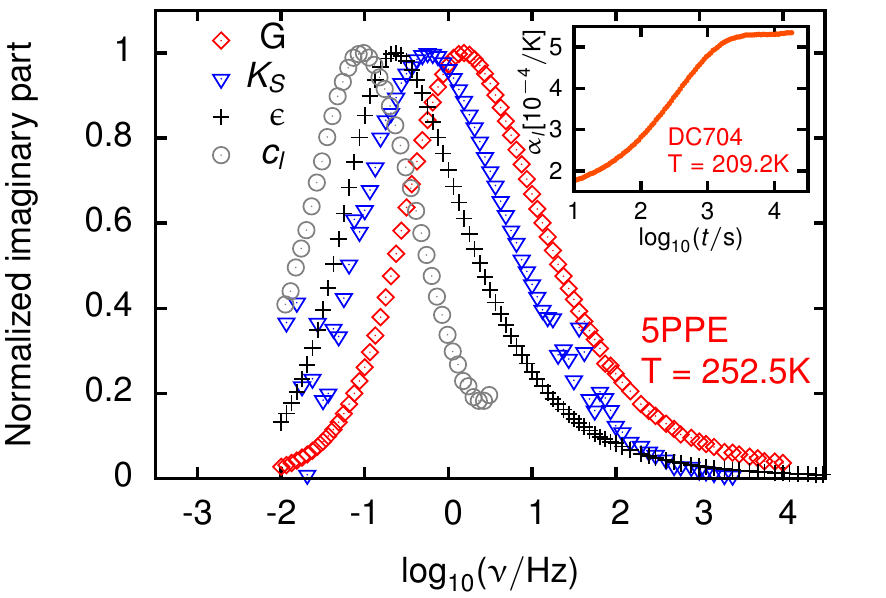}
  \caption{Example of data from which loss-peak positions are
    found. Main part: Normalized imaginary parts of the following
    complex response functions: shear modulus $G(\omega)$, adiabatic bulk
    modulus $K_S(\omega)$, dielectric constant
    $\epsilon(\omega)$, and  longitudinal specific heat $c_l(\omega)$,
    as functions of frequency for 5-polyphenyl-4-ether (5PPE) at
    $T=252.5\kelvin$. Inset: Time-dependent longitudinal expansion
    coefficient $\alpha_l(t)$ at
    $T=209.2\kelvin$ for tetramethyl-tetraphenyl-trisiloxane (DC704). 
    \label{fig:RawData}}
\end{figure}

Figure \ref{fig:RawData} shows the normalized loss peaks as functions
of frequency of $G(\omega)$, $K_S(\omega)$, $\epsilon(\omega)$, and
$c_l(\omega)$ for 5PPE at $252.5\kelvin$.  The inset shows one data
set for the time-dependent $\alpha_l(t)$ at $T=209.2\kelvin$ for
DC704. The frequency-domain data allow for direct determination of the
loss-peak frequencies ($\nu_\text{lp}$); the time-domain data were
Laplace transformed to give an equivalent loss-peak frequency
\cite{NissAlpha}.

\begin{figure}
  \centering
   \includegraphics[]{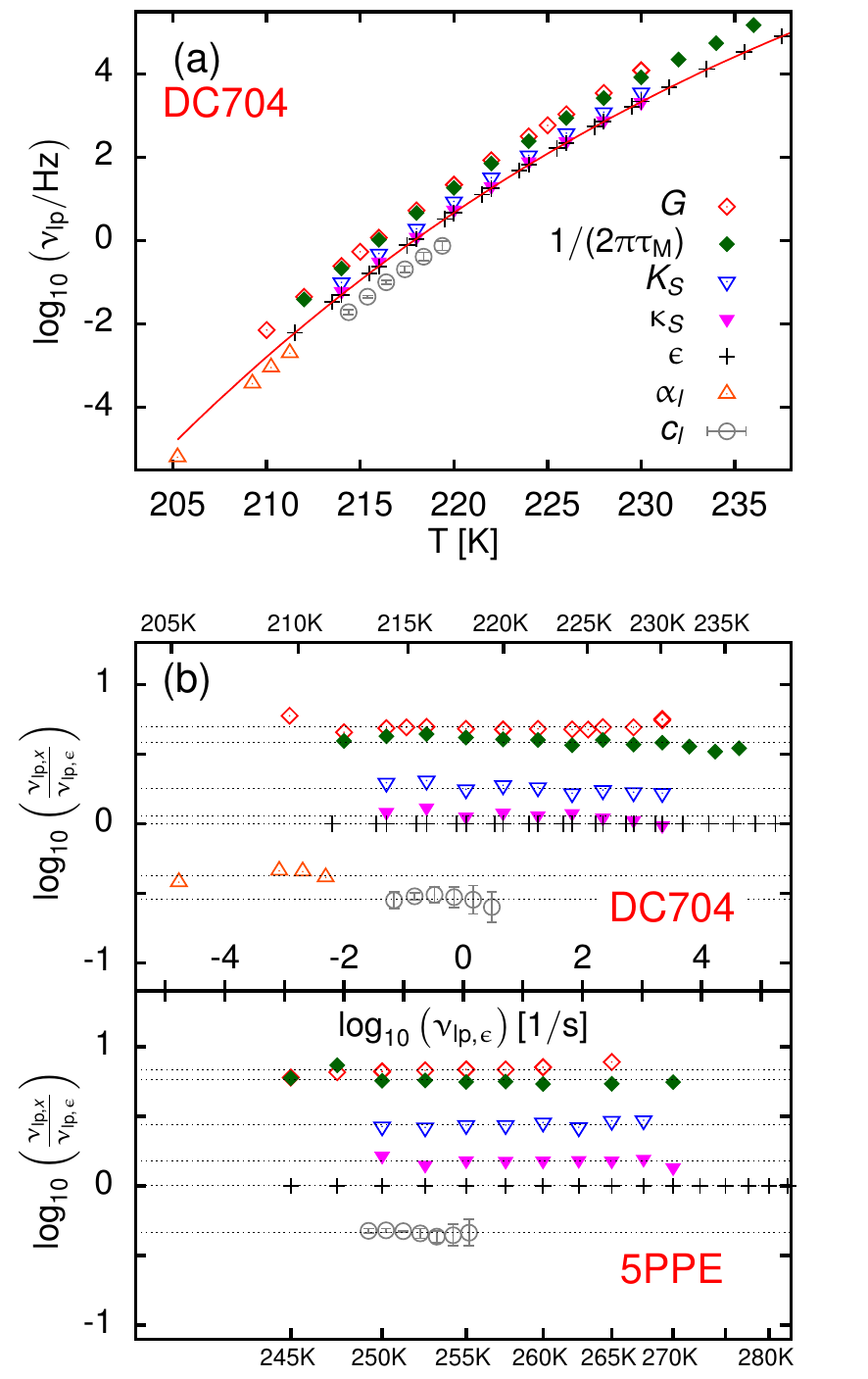}
   \caption{(a) Seven characteristic frequencies (based on five
     measured quantities) as functions of temperature for DC704. The
     full curve is a fit of the dielectric data to a Parabolic
     function\cite{Garrahan2003,*Elmatad2009,*Elmatad2010} (see
     appendix \ref{sec:choice-extr-funct} for details on this
     function), used for extrapolation of the
     dielectric data to low temperatures. Filled symbols
     ($\tau_\text{M}$ and $\kappa_S$) indicate that the quantity in
     question is not measured directly, but derived from one of the
     measured quantities. Error bars on $c_l$ data were estimated by
     assuming an additive influence from the underlying spurious
     frequency dependence of the raw data \cite{Jakobsen2010}, varying
     its influence on the loss peak from negligible to maximal.
     Equivalent data are given for 5PPE in appendix \ref{sec:5ppe-char-freq}. \\
     (b) Time-scale index of all measured response functions (symbols
     as in the top figure) with respect to the dielectric constant for
     the two liquids DC704 and 5PPE. The time-scale index is plotted
     as functions of the dielectric loss-peak frequency (common
     X-axis), which represents the temperature (also given for each
     liquid). For both liquids the time-scale index for all
     quantities are temperature independent within the experimental
     uncertainty, that is, the measured quantities have the same
     temperature dependence of their characteristic
     time scales. \label{fig:LPFandDec}}
\end{figure}

The measurements give both real and imaginary parts of the complex
response functions, allowing us to calculate two other relevant
characteristic frequencies, namely the inverse Maxwell time
$1/(2\pi\tau_\text{M})$ ($\eta$ and $G_\infty$ can be found from
$G(\omega)$) and the loss-peak frequency of the adiabatic
compressibility $\kappa_S(\omega)=1/K_S(\omega)$.  Figure
\ref{fig:LPFandDec}(a) shows the temperature dependence of the seven
characteristic frequencies for DC704\cite{online}.  The data
cover more than nine decades in relaxation time (from $T_g$ and
up). Clearly the time scales for the response functions follow each
other closely.

Figure \ref{fig:LPFandDec}(b) plots the characteristic frequencies in
terms of a \textit{``time-scale index''} defined as the logarithmic
distance to the dielectric loss-peak frequency\footnote{This
    quantity is sometimes refereed to as a ``decoupling index'';
    however, we reserve the word \textit{decoupling} for cases where the
    differences between the time scales are temperature dependent.}.
For both liquids the time-scale indices are temperature independent
within the experimental uncertainty, that is, the characteristic time
scales of the measured quantities change in the same way with
temperature. 

This finding constitutes the main result of the present paper, showing
that the time scales for theses response functions are strongly
coupled, in contrast to the observed decoupling between transitional
diffusion and rotation\cite{Fujara1992,Cicerone1996,Sillescu1999} and
at variance with the scenario suggested by Angell\cite{Angell1991}.

\section{Discussion}
The fastest response function is the shear modulus. It has previously
been reported that this quantity is faster than dielectric relaxation
for a number of glass-forming liquids (see e.g.\ Ref.\
\onlinecite{Deegan1999} and Ref.\ \onlinecite{Jakobsen2005} and
references therein), a fact that the Gemant-DiMarzio-Bishop model
explains qualitatively \cite{Niss2005}.  The dielectric relaxation is
faster than the specific heat (this difference cannot be attributed to
measuring $c_l$ and not $c_p$ \cite{supmat}).  For glycerol the same
tendency has been reported \cite{Ngai1990,Schroter2000}, but with a
fairly small difference in time scales of $c_p$ and $\epsilon$
($\approx 0.2$ decades). However a consistent interpretation of
dielectric hole-burning experiments on glycerol was arrived at by
assuming that these two time scales coincide \cite{Weinstein2005}. For
propylene glycol the opposite trend has been reported \cite{Ngai1990}.
Regarding volume and enthalpy relaxation there is likewise no general
trend in the literature; some glass-formers have slower enthalpy than
volume relaxation, others the opposite
\cite{Sasabe1978,Adachi1982,Badrinarayanan2007}. Clearly more
  work is needed to identify any possible general trends. However,
  such comparisons are difficult, as it requires a precision in
  absolute temperature at least better than $1\kelvin$ between the
  experiments. This is very difficult to obtain, and it could be
  speculated that some of the contradicting results could be explained
  this way. An advantage of our methods is that the same cryostat can
  be used for all the measurements ensuring same absolute temperature
  (see appendix \ref{sec:mater-meth-sect}).

What does theory have to say about the decoupling among relaxation
functions and why some are faster than others?  As mentioned, there
are three independent scalar thermoviscoelastic response
functions. There is no {\it a priori} reason these should have even
comparable loss-peak frequencies. Moreover, both the dielectric
constant and the shear modulus are linear-response functions that do
not belong to the class of scalar thermoviscoelastic response
functions; these two functions could in principle have relaxation
times entirely unrelated to those of the scalar thermoviscoelastic
response functions. All in all, general theory does not explain our
findings.

As mentioned earlier, the two investigated liquids obey TTS to a
  good approximation and have very small (if existing) beta
  relaxations. In an earlier work (Ref.\ \onlinecite{Jakobsen2005})
  some of us noticed that the time-scale index between shear
  mechanical and dielectric relaxation is only significantly
  temperature dependent for systems with a significant beta
  relaxation. In Ref.\ \onlinecite{Hecksher2010} it was shown for a
  number of systems (including systems with a pronounced beta
  relaxation) that the dielectric relaxation and the aging rate
  after a temperature jump follow the same \textit{``inner clock''};
  these results were obtained at temperatures where the alpha and beta
  relaxations are well separated.  Based on this, one might speculate
  that some (or all) of the temperature dependencies of the time-scale
  index observed in the literature could be due to the difference in
  the influence from the beta relaxation between the measured response
  functions (see e.g.\ Ref.\ \onlinecite{Jakobsen2011} for a
  comparison of the influence in shear mechanical and dielectric relaxation).

\subsection{Comparison to a ``perfectly correlating liquid''}
The class of so-called ``strongly correlating'' liquids was recently
identified
\cite{Pedersen2008,*Bailey2008,*Bailey2008a,*Schroder2009a,*Gnan2009,
  *Schroder2009,*Pedersen2010,*Gnan2010}. This class includes most or
all van der Waals and metallic liquids, but not the covalently-bonded,
hydrogen-bonded, or ionic liquids. In computer simulations strongly
correlating liquids are characterized by strong correlations between
constant-volume equilibrium fluctuations of virial and potential
energy
\cite{Pedersen2008,*Bailey2008,*Bailey2008a,*Schroder2009a,*Gnan2009,
  *Schroder2009,*Pedersen2010,*Gnan2010}. These liquids are
approximate single-parameter liquids, i.e., they do not have three
independent isotropic scalar thermoviscoelastic response functions,
but to a good approximation merely one \cite{Bailey2008b}. A perfect
single-parameter liquid obeys\cite{Ellegaard2007}
\begin{eqnarray}
  \label{eq:singelparm}
  \frac{T_0\alpha_p''(\omega)}{c_p''(\omega)}=\gamma_{Tp}=\frac{\kappa_T''(\omega)}{\alpha_p''(\omega)}
\end{eqnarray}
where $\gamma_{Tp}$ is a constant. Experimental evidence that DC704
is a strongly correlating liquid, was very recently presented in Ref.\
\onlinecite{Gundermann2011}.

We tested how well our results conform to the predictions for a
perfectly correlating liquid. If the thermoviscoelastic response
functions obey Eq.\ (\ref{eq:singelparm}), the loss-peak frequencies
should be identical at all temperatures. A single-parameter model
liquid was constructed by assuming Eq.\ (\ref{eq:singelparm}) to hold,
with high-frequency limits and relaxation strengths of the quantities
chosen to be as close as possible to those measured for DC704. The
quantities $\alpha_l(\omega)$, $c_l(\omega)$, and $\kappa_S(\omega)$
were calculated in the model by introducing shear-modulus data (see
appendix \ref{sec:perf-corr-single} for details).

\begin{figure}
  \centering
  \includegraphics[]{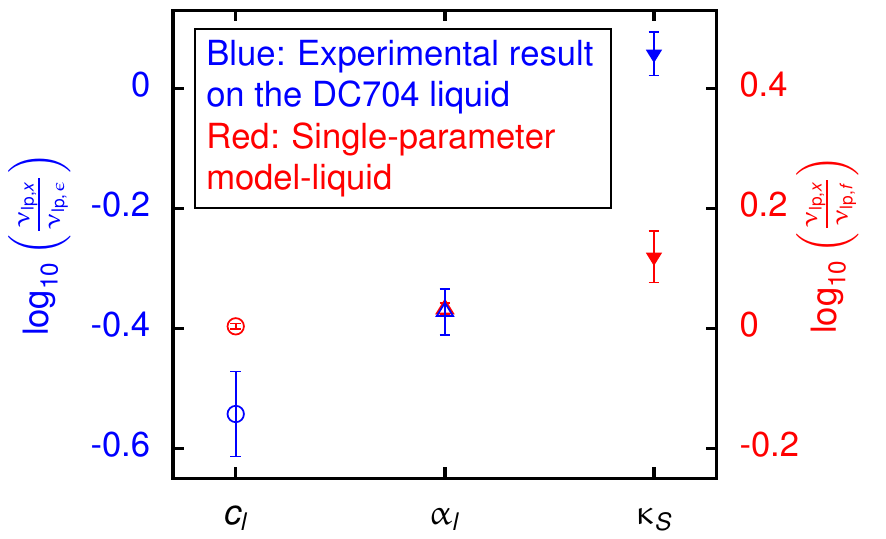}
  \caption{Time-scale indices of $c_l$, $\alpha_l$, and $\kappa_S$ for
    DC704 (blue symbols, left y-axis) and for the single-parameter
    model-liquid (red symbols, right y-axis) (see appendix
    \ref{sec:perf-corr-single} for details on the model
    calculation). The two axis have been shifted with respect to each
    other to give an overall best overlap of the indices.  For the
    experimental data the time-scale index is given relative to the
    dielectric loss-peak frequency ($\nu_{\text{lp},\epsilon}$). The
    reported quantity is the mean over the index in Fig.\
    \ref{fig:LPFandDec}(b), the error is based on the standard
    deviation (in the case of $c_l$ a mean over maximum and minimum
    based on the error bars on the loss-peak frequency was used). For
    the single-parameter model-liquid the index is with respect to the
    common loss-peak frequency of $\alpha_p$, $c_p$, and $\kappa_T$
    ($\nu_{\text{lp},f}$). The model liquid has the same order of the
    time scales as the real liquid, but the differences in the
    time scales are underestimated in the model. \label{fig:SCL}}
\end{figure}

Figure \ref{fig:SCL} compares the characteristic frequencies of $c_l$,
$\alpha_l$, and $\kappa_S$ of our experiments to those of the model
liquid. The order of the time scales for the model liquid matches the
observations, i.e., $c_l$ slower that $\alpha_l$ and $\kappa_S$ faster
than $\alpha_l$. However, the magnitude of the time-scale differences is
significantly underestimated.

\section{Summary}
We measured several complex frequency-dependent linear-response
functions on the two van der Waals liquids DC704 and 5PPE. Within the
experimental uncertainties the time scales of the response functions
have the same temperature dependence, that is, the \textit{time-scale
  indices} are temperature independent. The time scales are for both
liquids ordered from fastest to slowest as follows: Shear modulus,
adiabatic bulk modulus, dielectric constant, longitudinal thermal expansion
coefficient, longitudinal specific heat. 

General theory does not explain why the time scales from some response
functions couple very closely to each other, as is the case for the
investigated response functions, and why others show a decoupling, as
is observed, e.g.\ between transitional diffusion and
rotation\cite{Fujara1992,Cicerone1996,Sillescu1999}. The ordering of
the longitudinal thermal expansion coefficient, the longitudinal
specific heat and the adiabatic compressibility can be rationalized by
assuming that the liquids are strongly correlating, i.e., approximate
single-parameter
liquids\cite{Pedersen2008,*Bailey2008,*Bailey2008a,*Schroder2009a,*Gnan2009,
  *Schroder2009,*Pedersen2010,*Gnan2010,Ellegaard2007}, in which
certain sets of isotropic scalar thermoviscoelastic response functions
have identical time scales.

More work is indeed needed to establish if the findings are general
for van der Waals liquids, and to understand which response functions
have temperature independent time scale index and which decouple.

\begin{acknowledgments}
The centre for viscous liquid dynamics ‘‘Glass and Time’’ is sponsored
by the Danish National Research Foundation (DNRF).
\end{acknowledgments}


%

\clearpage

\appendix

\section{5PPE characteristic frequencies }\label{sec:5ppe-char-freq}

\begin{figure}
  \centering
  \includegraphics[]{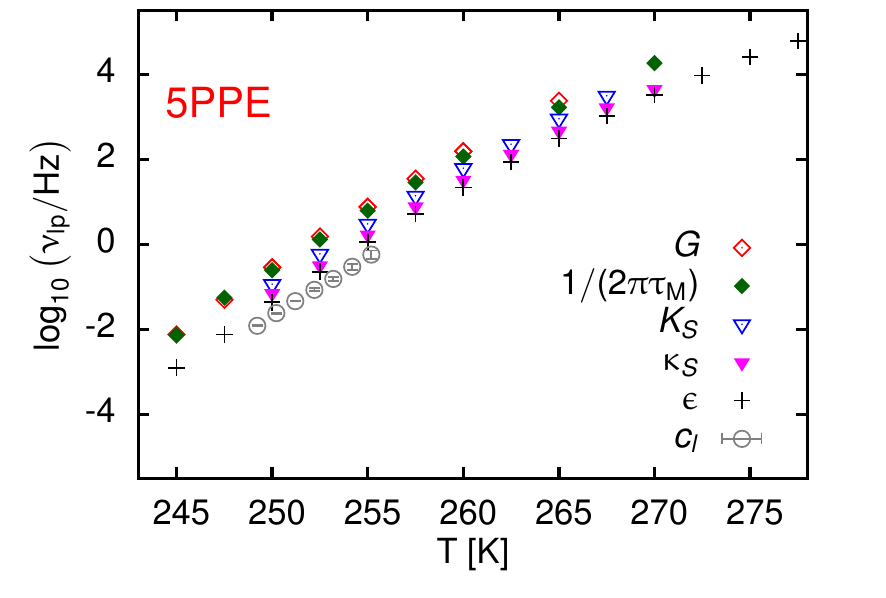}
  \caption{The six characteristic frequencies (based on the four
    measured quantities) as functions of temperature for 5PPE.  Filled
    symbols ($\tau_\text{M}$ and $\kappa_S$) indicate a
    quantity that is not directly measured, but derived from one of the
    other measured quantities.}
\vspace{2cm}
  \label{fig:Sup5PPElpf}
\end{figure}

Figure \ref{fig:Sup5PPElpf} shows the characteristic frequencies as
functions of temperature \cite{online} for 5-polyphenyl-4-ether (5PPE)
(the figure is equivalent to Fig.\ 2(a) in the main paper). Note that
$\alpha_l$ data are not available for 5PPE due to its larger
dielectric relaxation strength \cite{NissAlpha}.

\section{Materials and methods}
\label{sec:mater-meth-sect}
The two substances investigated are both commercial vacuum-pump oils
and were used as received without further purification. 
DC704 (tetramethyl-tetraphenyl-trisiloxane) was required from Aldrich
as Dow Corning \textregistered{} silicone diffusion pump fluid [445975].
5PPE (5-polyphenyl-4-ether) was required as Santovac \textregistered{}
5 Polyphenyl ether. 

With the exception of $\alpha_l(t)$ all quantities were measured in
the same cryostat \cite{Igarashi2008a,Igarashi2008b}, ensuring the
same absolute temperature. The thermal expansion coefficient data were
taken in a different cryostat, for which the absolute temperature was
calibrated using the liquid's dielectric relaxation time in a
temperature range where data exists from both cryostats. It was not
possible to measure $\alpha_l$ for 5PPE, because $\alpha_l$ was
measured by a method that requires a very small dielectric relaxation
strength \cite{NissAlpha}.

\section{The perfectly correlating (single-parameter) model liquid}
\label{sec:perf-corr-single}
As stated in the main text a strongly correlating liquid
\cite{Pedersen2008,*Bailey2008,*Bailey2008a,*Schroder2009a,*Gnan2009,
  *Schroder2009,*Pedersen2010,*Gnan2010}
is characterized in computer
simulations by strong correlations between the constant-volume
equilibrium fluctuations of virial and potential energy (correlation
coefficient above 0.9). Such a liquid is an approximate
single-parameter liquid, i.e., it does not have three independent
isotropic thermoviscoelastic response functions, but merely one (to a
good approximation)\cite{Ellegaard2007}.

The purpose of the model calculation is to investigate to which extent
our findings regarding the difference in characteristic frequency
between the thermoviscoelastic response functions $c_l$,
$\alpha_l$, and $\kappa_S$ are consistent with the hypothesis that the
investigated liquid tetramethyl-tetraphenyl-trisiloxane (DC704) is 
a strongly correlating liquid. For this purpose a perfect
single-parameter model liquid is constructed and the loss-peak
frequencies of the measured response functions are calculated.

At all frequencies a perfect single-parameter liquid obeys\cite{Ellegaard2007}
\begin{eqnarray}
  \label{eq:SupSubgekParm}
  \frac{T_0\alpha_p''(\omega)}{c_p''(\omega)}=\gamma_{Tp}=\frac{\kappa_T''(\omega)}{\alpha_p''(\omega)}
\end{eqnarray}
where $\gamma_{Tp}$ is a constant. By the Kramers-Kronig relations
Eq.\ (\ref{eq:SupSubgekParm}) implies that the relaxation strengths
are related by
\begin{eqnarray}
  \label{eqSupDelta}
  \frac{T_0\Delta{\alpha_p}}{\Delta{c_p}}=\gamma_{Tp}=\frac{\Delta{\kappa_T}}{\Delta{\alpha_p}}.
\end{eqnarray}

It follows that the full response functions can be written as
\begin{subequations} \label{eq:SupSCL}
  \begin{eqnarray}
    \alpha_p(\omega)&=&\Delta{\alpha_p}f(\omega)+\alpha_{p\infty}\\
    c_p(\omega)&=&\frac{T_0}{\gamma_{Tp}}\Delta{\alpha_p}f(\omega)+c_{p\infty}\\
    \kappa_T(\omega)&=&\gamma_{Tp}\Delta{\alpha_p}f(\omega)+\kappa_{T\infty},
  \end{eqnarray}
\end{subequations}
where $f(\omega)$ is a normalized complex susceptibility, defining the
common time scale as well as the shape of the relaxation
functions. $\alpha_{p\infty}$, $c_{p\infty}$, and $\kappa_{T\infty}$
are the high-frequency limits of the response functions, and
$\Delta{\alpha_p}$ is the relaxation strength of $\alpha_p$.

The measured quantities $c_l$, $\alpha_l$, and $\kappa_S$ are related
to those of  Eq.\ (\ref{eq:SupSCL}) by
\cite{Christensen2007,Christensen2008}:
\begin{subequations} \label{eq:SupMeasuredQuantities}
  \begin{eqnarray}
    \alpha_l(\omega)&=&\frac{\alpha_p(\omega)}{1+4/3\; G(\omega)\kappa_T(\omega)}\\
    c_V(\omega)&=&c_p(\omega)-\frac{T_0\alpha_p(\omega)^2}{\kappa_T(\omega)}\\
    \kappa_S(\omega)&=&\frac{c_V(\omega)}{c_p(\omega)}\kappa_T(\omega)\\
    c_l(\omega)&=&\frac{1/\kappa_S(\omega)+4/3\;  G(\omega)}{1/\kappa_T(\omega)+4/3\; G(\omega)}c_V(\omega).
  \end{eqnarray}
 \end{subequations}
Thus if the shear modulus is known, it is possible to
calculate the measured quantities based on the quantities which are
related in the single-parameter liquid: $\alpha_p$, $c_p$, and
$\kappa_T$.

\begin{table}[b]
  \caption{Parameters for the perfect single-parameter liquid mimicking the
    properties of DC704 at the temperature  $214\kelvin$. The values
    are based on the data presented in Refs.\
    \onlinecite{Gundermann2011,HecksherPHD}.\label{tab:SPModel} }
  \centering
\renewcommand{\arraystretch}{1.3}
  \begin{tabular}{lcl}  
    \hline   \hline
    $\gamma_{Tp}$&:& $2.5\E{-7}\meter^3\kelvin\per\joule$ \\ 
    $\Delta{\alpha_p}$ &:&$3.2\E{-4} 1/\kelvin$\\
    $\alpha_{p\infty}$ &:&$1.4\E{-4} 1/\kelvin$ \\
    $c_{p\infty}$&:& $1.4\E{6}\joule\per(\meter^3\kelvin)$\\ 
    $\kappa_{T\infty}$&:&$1.9\E{-10}\meter^3\per\joule$\\ \hline
    $G_{\infty}$&:& $3/5\Delta{K_S}$\\
    $\log_{10}\left(\frac{\nu_{\text{lp},G}}{\nu_{\text{lp},f}}\right)$&:&1.1\\
    \hline   
    $f(\omega)$&:& $\left({1+\frac{1}{(i\omega\tau)^{-1} +
          k(i\omega\tau)^{-\alpha}}}\right)^{-1}$\\
    $\alpha$ &:& 0.44 \\
    $k$&:& $1.05$ \\
    \hline \hline
  \end{tabular}   
\end{table}

 \begin{figure}[b]
   \centering
   \includegraphics[]{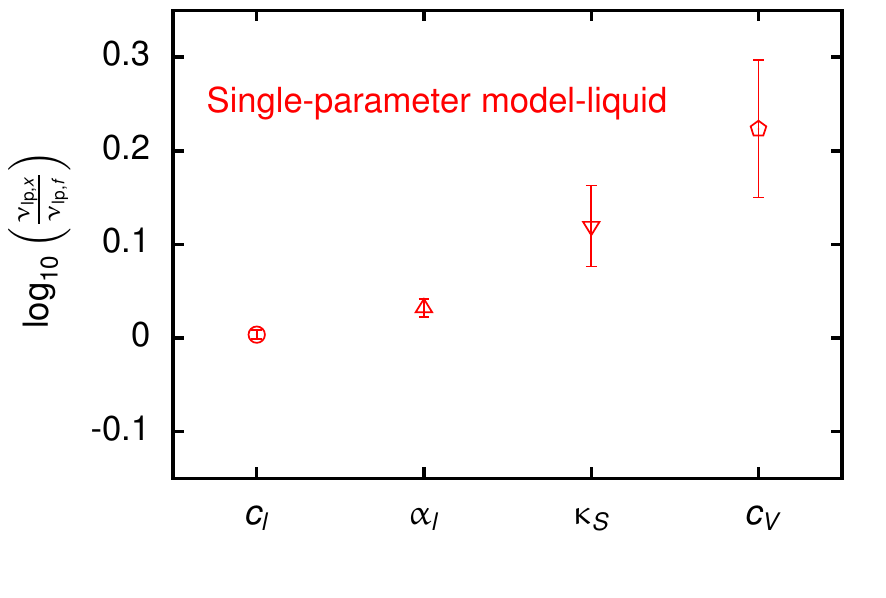}
   \caption{Time-scale index for $c_l$, $\alpha_l$, $\kappa_S$, and
     $c_V$ relative to the loss peak position of the common function
     $f(\omega)$ entering Eq.\ (\ref{eq:SupSCL}).  \label{fig:SupDec}}
 \end{figure}

A single-parameter model liquid with properties close to DC704 was
constructed in the following way (the parameters used are listed in
Table \ref{tab:SPModel}) :
\begin{itemize}
\item The high-frequency limits and relaxation strengths of
  $\alpha_p$, $c_p$, and $\kappa_T$ were calculated based on the high-
  and low-frequency limits of of $c_l$, $\kappa_S$, $\alpha_l$, and
  $G$ as given in Ref.\ \onlinecite{Gundermann2011}. This can be done,
  because Eq.\ (\ref{eq:SupMeasuredQuantities}) can be inverted
  analytically.
\item $\gamma_{Tp}$ was estimated from Eq. (\ref{eqSupDelta}).  DC704 is
  not a perfectly correlating liquid, so the right- and
  left-hand sides do not give exactly the same value. They differ by
  $\approx 10\%$, which is below their respectively uncertainties, and
  the average was used.
\item The normalized complex susceptibility $f(\omega)$ was chosen to
  follow a ``generalized BEL model'', which we found to describe
  the alpha relaxation in many liquids very well
  \cite{Christensen1994b,Saglanmak2010,Jakobsen2011}:
  \begin{eqnarray}
    \label{eq:Supf}
    f(\omega)=\frac{1}{1+\frac{1}{(i\omega\tau)^{-1} +
        k(i\omega\tau)^{-\alpha}}}. 
  \end{eqnarray}
  The parameters were estimated by fitting a modulus version of the
  model to the shear modulus data.
\item The shear modulus was assumed to follow a modulus version of
  Eq.\ (\ref{eq:Supf}), having $G_{\infty}=3/5\Delta{K_S}$ (in Ref.
  \onlinecite{HecksherPHD} this relation was shown to apply for DC704),
  and a time scale that is one decade faster than $f(\omega)$ (as
  seen in Fig.\ 2 of the main paper).
\item $10^5$ realizations of the model liquid were calculated
  with the parameters $\gamma_{Tp}$, $\Delta{\alpha_p}$,
  $\alpha_{p\infty}$, $c_{p\infty}$, and $\kappa_{T\infty}$ chosen
  independently from Gaussian distributions with mean given as in Table
  \ref{tab:SPModel} and a standard deviation of $20\%$.  This was done
  in order to investigate the robustness of the results with respect
  to the influence of the absolute levels. The chosen standard
  deviation is larger than the estimated uncertainties on the
  quantities, ensuring that the results represent a ``worst
  case scenario''. For each realization the loss-peak positions of $c_l$,
  $\alpha_l$, and $\kappa_S$ (and additionally $c_V$) were calculated.
\end{itemize}

Figure \ref{fig:SupDec} shows the time-scale index between the
loss-peak frequencies of the calculated quantities with respect to the
loss-peak frequency of $f(\omega)$ (Eq.\ (\ref{eq:Supf})). That is,
Fig.\  \ref{fig:SupDec} shows the loss-peak positions of $c_l$, $c_V$,
$\alpha_l$ and $\kappa_S$ when the underlying liquid is a
perfect single-parameter liquid in which $c_p$, $\alpha_p$, and
$\kappa_T$ have the same loss-peak position. It can be seen that the
influence on the loss-peak position of measuring the longitudinal
specific heat and thermal expansion coefficient, instead of their
isobaric counterparts, is very small. Compressibility, on the other
hand, is affected more by measuring the adiabatic version instead
of the isothermal.

\section{Choice of extrapolation function}\label{sec:choice-extr-funct}
\begin{figure}
  \centering
   \includegraphics[]{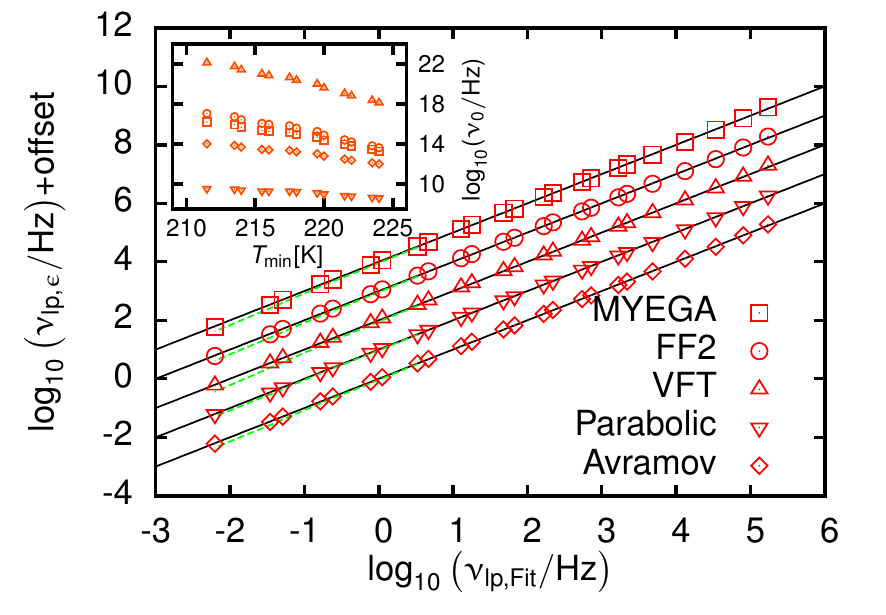}
   \caption{Quality of five fitting functions of the loss-peak
     frequency as functions of temperature. The main figure shows the
     result of evaluating fits of the five different function to the
     full data set (see Table \ref{tab:Fitfkt} for details on
     expressions and parameters and the main text for references),
     plotted against the measured loss-peak frequencies (for clarity
     the plots have been offset by an integer constant on the
     y-axis). For a perfect fit the points fall on the shown black
     lines. The (barely visible) dashed green lines show the ability
     to extrapolate; it is the result of evaluating a fit including
     only data down to $\log_{10}(\nu_\text{lp})=0.67$ (corresponding
     to $220\kelvin$). The insert shows the best fit attack
     frequencies, $\nu_0$, as functions of the lowest temperature
     included in the fit (see the main text for details). This figures
     shows that all five functions fit data well, but the Parabolic
     function is the best function to used for extrapolation of data
     to lower temperatures.
     \label{fig:Fitfkt}}
\end{figure}

\begin{table*}
\renewcommand{\arraystretch}{2}
  \centering
  \caption{Fitting functions. The table
    presents the five functions (see main text for references) which
    were tested for the ability 
    to extrapolate the dielectric loss peak to the temperatures where the
    thermal expansion coefficient was measured. The parameters
    reported derive from fitting the full data set. The ``Parameter
    change'' is the 
    relative change in 
    the parameters between fitting to the 12 highest temperatures
    (down to $224\kelvin$) and
    to the full set of 24 points (down to
    $211.5\kelvin$). \label{tab:Fitfkt}}   
  \begin{minipage}{\linewidth}  
    \begin{tabular}{llll}
      \hline \hline
      & $\nu_\text{lp}(T)$ & Parameters & Parameter chage\\ \hline
      VFT &
      {$\nu_0\exp\left(-\frac{A}{T-T_0} \right)$}&
      $A=3510\kelvin, T_0=148.9\kelvin,\log_{10}(\nu_0)=22.1$ &
      $A:62\% , T_0:-10\%,\log_{10}(\nu_0):22\%$
      \\
      Avramov 
      &
      $\nu_0\exp\left(-\left(\frac{B}{T}\right)^n\right)$&
      $n=5.0,B=437.0\kelvin,\log_{10}(\nu_0)=14.0$&
      $n:-19\%,B:17\%,\log_{10}(\nu_0):17\%$
      \\
      Parabolic &
      {$\nu_0\exp\left(-J^2\left(\frac{1}{T}-\frac{1}{T_0}\right)^2\right)$}&
      $J=3740\kelvin^2,T_0=299.6\kelvin,\log_{10}(\nu_0)=9.5$&
      $J:-8\% ,T_0:5\% ,\log_{10}(\nu_0):11\% $
      \\
      Double exp (MYEGA) &
      $\nu_0\exp\left(-\frac{K}{T}\exp\left(\frac{C}{T}\right)\right)$&
      $K=294.4\kelvin, C=722.5\kelvin, \log_{10}(\nu_0)=16.2$&
      $K:344\%, C:-28\%, \log_{10}(\nu_0):22\%$
      \\
      Double exp (FF2)&
      $\nu_0\exp\left(-A\exp\left(\frac{T_2}{T}\right)\right)$&
      $A=0.650, T_2=893.5\kelvin,\log_{10}(\nu_0)=17.0$&
      $A:393\%, T_2:-25\% ,\log_{10}(\nu_0):25\%  $

      \\
      \hline \hline
    \end{tabular}
  \end{minipage}
\end{table*}

In order to extrapolate the dielectric data to the temperatures of the
thermal expansion coefficient, a fitting function is needed. Inspired
by the work of Elmatad \textit{et al.}\ \cite{Elmatad2010} five
fitting functions were investigated (Table \ref{tab:Fitfkt}): The
Vogel-Fulcher-Tammann (VFT) function
\cite{Vogel1921,Fulcher1925,Tammann1925}, the Avramov function
\cite{Litovitz1952,Barlow1959,Barlow1966,Harrison1976,Bassler1987,Avramov2005},
the Parabolic function \cite{Garrahan2003,Elmatad2009,Elmatad2010},
and two double-exponential functions: The MYEGA function suggested by
Mauro \textit{et at.}  \cite{Mauro2009} and the similar FF2 function
of Hecksher \textit{et al.}  \cite{Hecksher2008}. All five functions
have three fitting parameters, one of which is the microscopic attack
frequency ($\nu_0$).

Each function was fitted to the loss-peak position from the dielectric
constant, using a least-square minimization. Figure \ref{fig:Fitfkt}
illustrates the quality of the fits, which for all five functions are
excellent. The fitting parameters are given in Table
\ref{tab:Fitfkt}. In order to explore to which extent the functions
are useful for extrapolating, the stability of the parameters were
investigated when adding points to the fit. All functions were fitted
to the 12 points at the highest temperatures. Points were then added
to the data set one by one and the functions re-fitted. The relative
change in the fitting parameters from the initial to the final data
set is reported in Table \ref{tab:Fitfkt}, and the insert in Fig.\
\ref{fig:Fitfkt} shows the fits for the attack frequencies ($\nu_0$).
A clear convergence of the parameters was not observed for any of the
functions, but the Parabolic function has the most stable parameters
(however, the Avramov function is only slightly worse and it has a
more physically realistic attack frequency).  Finally, the ability to
extrapolate was investigated by fitting to the first 16 points and
then evaluating the fitted function at the lower temperatures, as
shown by the dashed green lines in Fig.\ \ref{fig:Fitfkt}). The
extrapolation spans $\approx 2.5$ decade in loss-peak frequency,
comparable to the extrapolation needed to get to the lowest
temperature of the thermal expansion coefficient. Again the Parabolic
function is best, but not perfect. Based on this the Parabolic function
was chosen as extrapolation function.

\end{document}